# Quantum Hall Fabry–Pérot Interferometry in Monolayer $WSe_2$

Himadri Chakraborti[1,2†*], Charles L. Tschirhart[2,3,4†*], James Fang[2], Saisab Bhowmik[1], Chengrui Yu[2], Shujian Cao[2], William Caro[2], Kenji Watanabe[5], Takashi Taniguchi[6], Jie Shan[1−4], Kin Fai Mak[1−4]

[1] Max Planck Institute for the Structure and Dynamics of Matter, Hamburg, Germany
[2]Laboratory of Atomic and Solid State Physics, Cornell University, Ithaca, NY 14853, USA
[3]School of Applied and Engineering Physics, Cornell University, Ithaca, NY 14853, USA
[4]Kavli Institute at Cornell for Nanoscale Science, Cornell University, Ithaca, NY 14853, USA
[5]Research Center for Electronic and Optical Materials, National Institute for Materials Science, Tsukuba, Japan
[6] Research Center for Materials Nanoarchitectonics, National Institute for Materials Science, Tsukuba, Japan
[†]These authors contributed equally.
[*]Correspondence: himadri.chakraborti@mpsd.mpg.de, cltschirhart@gmail.com

## Abstract

**Electronic Fabry–Pérot interferometers provide a powerful probe of coherent quantum Hall edge transport, yet the presence of multiple spin and valley degrees of freedom in existing van der Waals material platforms can complicate their interpretation. Monolayer $WSe_2$ offers a distinct opportunity because strong spin–orbit coupling and spin–valley locking lift these degeneracies, producing intrinsically single-component Landau levels. Here we realize the first gate-defined quantum Hall Fabry–Pérot interferometer in monolayer $WSe_2$. By continuously tuning the cavity size in a single device, we observe robust Aharonov–Bohm (AB) interference of quantum Hall edge channels across the interferometer; the observed predominance in the AB regime is consistent with a weak bulk–edge coupling ($\Delta \approx 0.034$) estimated by electrostatics. Moreover, temperature and bias dependence measurements consistently yield an edge velocity $v \approx (4.9 \pm 0.2) \times 10^4\ \mathrm{ms^{-1}}$. Our results establish monolayer $WSe_2$ as a clean, flavor-resolved platform for quantum Hall interferometry and provide a foundation for future studies of single-component fractional quantum Hall states and interferometric probes of anyonic statistics in transition-metal dichalcogenides.**

## Main

Electronic interferometers provide a direct probe of the quantum coherence and phase of charge carriers in low-dimensional systems. In the quantum Hall (QH) regime, where transport occurs through chiral edge channels, two geometries have emerged as the principal platforms for interferometry: the Mach–Zehnder interferometer, in which two edge channels are coupled by spatially separated quantum point contacts (QPCs), and the Fabry–Pérot (FP) interferometer, in which a single edge channel is partially backscattered at two QPCs enclosing a gate-defined cavity[1–8]. The FP geometry is particularly powerful because the interference phase directly encodes both the Aharonov–Bohm (AB) phase and, in the fractional QH regime, the statistical phase acquired by quasiparticles encircling the cavity[9–16]. This capability makes FP interferometry one of the leading approaches for probing fractional statistics and non-Abelian anyons[5,7,17–21]. Achieving this goal, however, requires compact cavities with long phase coherence, precise electrostatic control of the edge channels, and operation in the AB regime, where charging effects do not dominate the interference[9,22]. Simultaneously satisfying these conditions has remained a major experimental challenge.

Gate-defined QH FP interferometers have been realized in GaAs heterostructures[10–12,23–25] and, more recently, in graphene[14,26–28]. In both platforms, shrinking the cavity to improve the ratio of phase-coherence length to cavity perimeter generally enhances Coulomb charging effects, which can obscure the AB signal[9,29]. Graphene also retains four-fold spin–valley degeneracy[30], making the symmetry of the interfering channel ambiguous, particularly in the fractional regime where even-denominator states are expected to host non-Abelian quasiparticles[31].

Monolayer $WSe_2$ could overcome these limitations. Strong spin–orbit coupling and spin–valley locking lift the spin and valley degeneracies, yielding single-component Landau levels without requiring a displacement field or additional electrostatic control[32–35]. Fractional QH states, including even-denominator states in a single-component Landau level, have already been observed in this material[36], making it an attractive platform for interferometric studies of single-component fractional QH physics. At the same time, its atomically thin geometry enables sharp electrostatic confinement and compact gate-defined cavities, while its direct-gap band structure opens opportunities to combine electronic interferometry with optical probes of spin, valley, and excitonic degrees of freedom[37,38].

Despite these advantages, realizing a small-scale gate-defined quantum Hall interferometer working in the AB regime in monolayer $WSe_2$ has remained experimentally challenging. Doing so requires combining clean materials, low-resistance ohmic electrical contacts, precise electrostatic confinement, and reproducible control of edge-channel transmission in a platform where quantum Hall interferometry had not previously been demonstrated.

Here we demonstrate the first gate-defined QH FP interferometer in monolayer $WSe_2$, operating at an integer filling factor $\nu = -4$. The choice of $\nu = -4$ is primarily set by the maximum magnetic field (12 T) in our setup, which corresponds to a carrier density ($\approx 10^{12}$cm$^{-2}$ at $\nu = -4$) high enough for interferometer operation in this material. We interfere the outermost edge channel (Figure 1C), while the enclosed inner edge channels are expected to partially screen it from electrostatic fluctuations associated with the bulk, favoring robust phase-coherent interference[9,26,39]. By continuously tuning a single device through three cavity sizes, we realize compact interferometers with perimeters $2L \approx 740–880$ nm, among the smallest gate-defined QH FP cavities reported to date. All three exhibit clear AB interference. From their bias and temperature dependence we extract a common edge velocity, ballistic Thouless energy, and phase-coherence length. An independent electrostatics analysis further shows a weak bulk–edge coupling ($\Delta \approx 0.034$), confirming the operation of these compact cavities in the AB regime rather than the Coulomb-dominated regime. Together, these results establish monolayer $WSe_2$ as a promising platform for coherent QH interferometry and provide a route toward interferometric studies of single-component fractional QH states and non-Abelian quasiparticles.

## Interferometer Design

The FP interferometer is built from chiral QH edge channels in monolayer $WSe_2$, encapsulated between two hexagonal boron nitride (hBN) gate dielectric layers. A global graphite top gate tunes the carrier density of the $WSe_2$ layer. The interferometer is defined by two QPCs, which bring the counter-propagating edge channels into close proximity and provide controlled backscattering.

Two plunger-gates define the cavity and continuously tune its effective area, as illustrated in Figure 1.

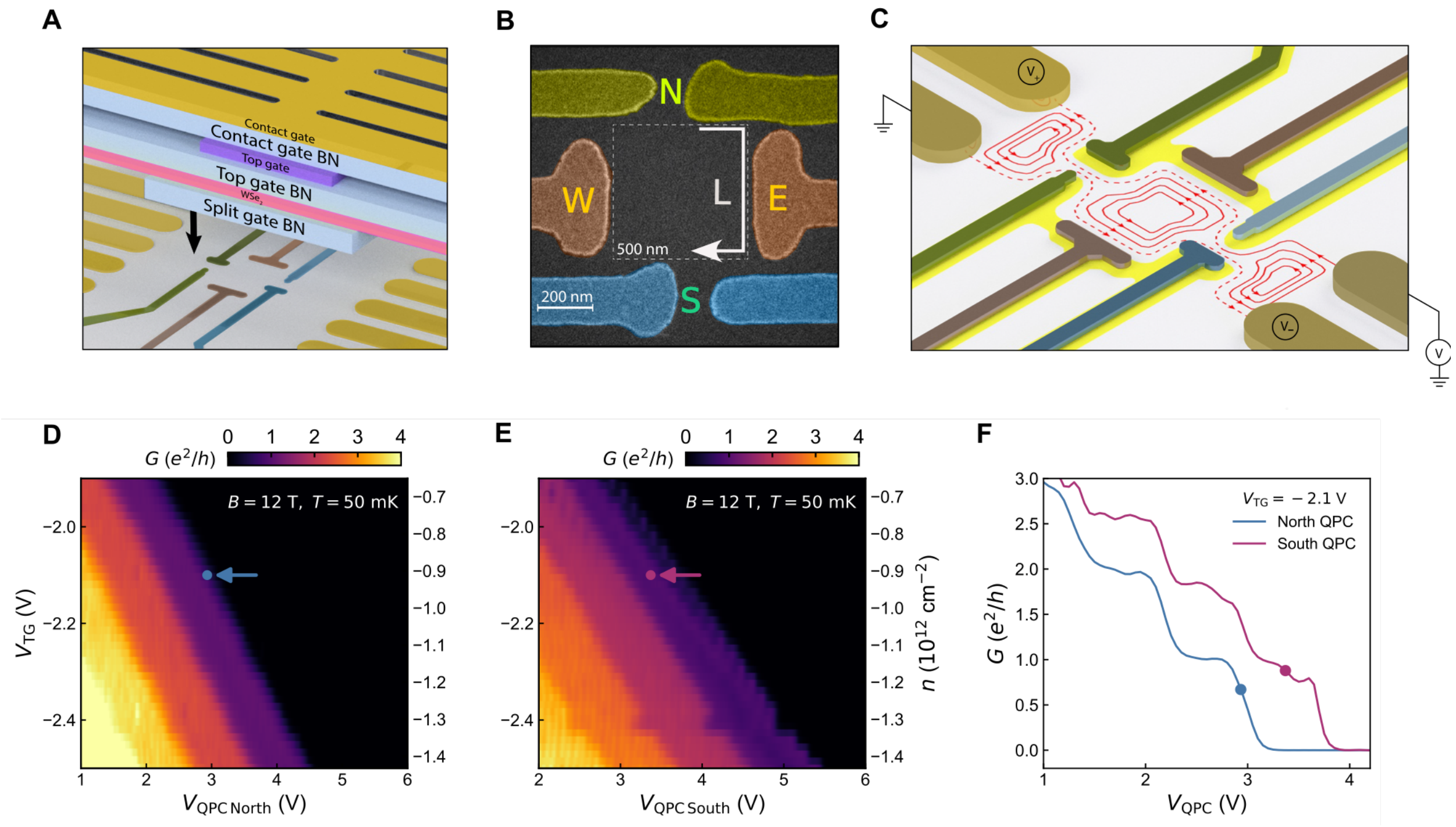

**Figure 1: Device architecture and characterization of the tunable Fabry–Pérot interferometer. A,** Schematic cross-section of the van der Waals heterostructure: a $WSe_2$ monolayer (light magenta) encapsulated between hBN dielectric layers (blue), with a graphite top gate (purple) and a Pd contact gate (golden yellow). The split-gate geometry defining the QPCs is visible beneath the stack. $WSe_2$ layer is in direct contact with the pre-patterned Pt electrodes (yellow). **B,** False color scanning electron micrograph of the metallic QPC- and plunger-gate electrodes pre-patterned on the Si/$SiO_2$ substrate. Scale bar: 200 nm. **C,** Top-view schematic of the Fabry–Pérot cavity showing the four-terminal measurement configuration in the quantum Hall regime at $\nu = -4$. Red dashed lines indicate the outermost interfering chiral edge-channel. **D,E,** Diagonal conductance as a function of the QPC-gate and top-gate voltages for the north (**D**) and south (**E**) QPC at $B = 12$ T and $T = 50$ mK, showing quantized edge-channel transmission in the quantum Hall regime. **F,** High resolution line cuts at $V_{TG} = -2.1$ V. The dot and arrow in **D** and **E** indicate the operating points used for the interferometry measurements.

The QPC- and plunger-gate electrodes are patterned on a Si/$SiO_2$ substrate by electron-beam lithography (Cr/Pd, 2/10 nm). The van der Waals heterostructure is then assembled and transferred onto the pre-patterned gates. Low-resistance ohmic contacts with hole-doped $WSe_2$ are made with pre-patterned platinum (Pt) electrodes and a metallic contact gate on top of the heterostructure to heavily hole-dope the contact regions and to turn on the device. The device has six usable contacts, three on each side of the interferometer (Figure S1).

Before performing the interferometry measurements, we first establish the transmission characteristics of each QPC under a perpendicular magnetic field $B = 12$ T and sample temperature $T \approx 50$ mK (the default setting unless otherwise specified). Figure 1D shows the diagonal conductance $G$ of the north QPC, measured between the $V_+$ and $V_-$ contacts (Figure 1C), as a function of the top-gate ($V_{\text{TG}}$) and QPC-gate ($V_{\text{QPC North}}$) voltages. The corresponding result for the south QPC is shown in Figure 1E. As expected, a more positive $V_{\text{QPC}}$ is required to pinch off the QPC for a higher hole-doping density (i.e. a more negative $V_{\text{TG}}$) in the monolayer $WSe_2$. Well-quantized conductance plateaus (in units of $\frac{e^2}{h}$, see line cuts in Figure 1F) are observed, illustrating the high quality of the gate-defined QPCs[40–42]. (Here $e$ and $h$ denote the electron charge and Planck's constant, respectively.)

The result above demonstrates the successive pinching of each pair of counter-propagating QH edge channels as the QPC constriction is progressively depleted; it also allows us to identify the appropriate operating points for controlled partitioning of the interfering edge channel at each QPC. In the following, we set $V_{\text{QPC}}$ near the $G \approx \frac{e^2}{h}$ plateau for each QPC (marked as filled circles and arrows in Figure 1F) and the bulk filling factor to $\nu = -4$ ($V_{\text{TG}} = -2.1$ V). It corresponds to the configuration illustrated in Figure 1C, where four QH edge channels propagate along the sample boundary with the outermost edge (dashed) being used for interference. The transmission of the outermost edge channel is about $T_N = 0.65$ and $T_S = 0.91$ for the north and south QPC, respectively.

**Aharonov–Bohm oscillations at $\nu = -4$**

Figure 2A shows the diagonal conductance $G$ as a function of the east plunger-gate voltage, with the west plunger-gate held fixed at 2.0 V. Unless otherwise specified, all measurements in this paper are performed in this configuration. From this point on we drop the east/west distinction and refer to the varied plunger-gate simply as the plunger-gate, $V_{\text{PG}}$. Clear periodic oscillations are visible across the full plunger-gate range; the oscillation period increases with increasing $V_{\text{PG}}$. By operating $V_{\text{PG}}$ at three different working windows (colored), we realize three different effective cavity areas in the same device, hereafter referred to as the large, medium, and small configurations. Their distinct areas are later confirmed by their different magnetic-field periods. Figures 2B–D show magnified traces of the conductance oscillations within each window. The oscillations are well resolved, with comparable peak-to-trough visibilities $(G_{\max} - G_{\min})/(G_{\max} + G_{\min}) \approx$ 6.3%, 6%, and 8.2% for the large, medium, and small configurations, respectively.

We further examine the diagonal conductance as a function of $V_{\text{PG}}$ and $B$ for the three configurations in Figures 2E–G. The smooth evolution of the observed fringe pattern across the three configurations confirms that the plunger-gate provides continuous electrostatic control over the cavity. The fringes follow the constant-flux condition $\Phi = BA_{\text{eff}} = \text{const}$, i.e. an increase in $B$ must be compensated by a reduction of the effective area $A_{\text{eff}}$, achieved by increasing $V_{\text{PG}}$, to keep the oscillation phase fixed. The resulting positive fringe slope, $\text{d}B/\text{d}V_{\text{PG}} > 0$, is a hallmark of AB interference, confirming robust phase coherent transport throughout the tunable range of the interferometer. The result is distinct from Coulomb-dominated oscillations, which would show the opposite slope[9].

The observed magnetic-field periods $\Delta B \approx 86$, 100, and 120 mT correspond to effective areas $A_{\text{eff}} = \frac{h}{e\Delta B} \approx 0.048$, 0.041, and 0.034 $\mu\text{m}^2$ for the large, medium, and small configuration, respectively. The extracted $A_{\text{eff}}$ is substantially smaller than the lithographically defined cavity area ≈ 0.25 $\mu\text{m}^2$. The origin of this discrepancy is not fully understood, but it may arise from a narrow region of reduced carrier density near the cavity boundaries, which shifts the interfering edge channel inward, and/or local imperfections that modify the effective cavity geometry. A reduction in the effective conducting width relative to the lithographic dimensions is also observed in zero-field transport measurements (Supplementary Fig. S2).

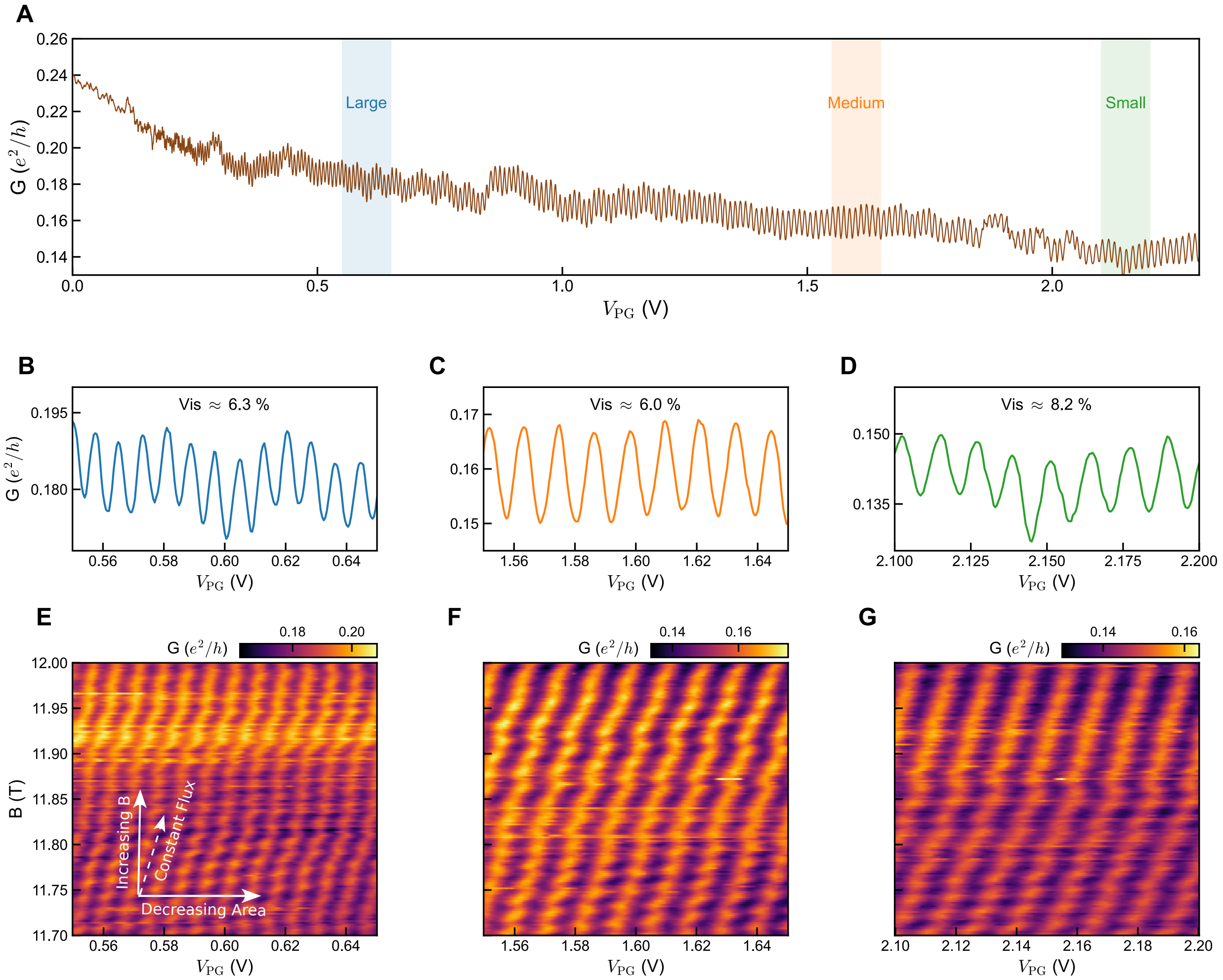

**Figure 2: Aharonov–Bohm interference at three cavity sizes of the tunable interferometer. A,** Diagonal conductance $G$ versus plunger-gate voltage $V_{PG}$, measured at $B = 12$ T, $T = 50$ mK and bulk filling factor $\nu = -4$. The transmissions of the north and south QPCs are set to $T_N \approx 0.65$ and $T_S \approx 0.91$, respectively. Periodic oscillations are resolved across the full gate range; the three colored windows mark the cavity sizes selected for analysis. **B–D,** Magnified conductance traces within the three windows (large, medium, small), showing well-resolved AB oscillations. **E–G,** $G$ versus $V_{PG}$ and $B$ for the three cavity sizes. The interference forms diagonal stripes of constant enclosed flux, with the fringe slope annotated in **E**.

While the observed fringes with positive slopes confirm the AB-dominated operation in our device, the presence of additional periodic modulations on top of the fringes indicate the non-negligible

Coulomb interaction effects[9]. To quantify the relative importance of Coulomb interactions, we estimate the dimensionless bulk–edge coupling parameter $\Delta = C_{IL}/(C_{IL} + C_L) \approx 0.034$[9,43] from the device electrostatics (Supplementary Section S3). (Here $C_L$ and $C_{IL}$ denote the bulk and the edge-to-bulk capacitances of the interferometer, respectively.) The small $\Delta \approx 0.034$ indicates weak capacitive coupling between the interfering edge and the localized charges in the bulk, confirming the predominant operation in the AB regime (with small corrections from residual interaction effects). The value is also comparable to the reported one for AB-dominated graphene FP interferometers[26,27].

**Temperature dependence of the interference visibility**

Figures 3A–C show the diagonal conductance $G$ as a function of $V_{\mathrm{PG}}$ and $T$ for the three cavity sizes. The periodic AB oscillations gradually disappear with increasing temperature; only a smooth, gate-independent background remains at $T \gtrsim 0.6$ K. To quantify the temperature dependence, we extract at each temperature the Fourier amplitude of the conductance oscillation at the fringe frequency (in $V_{\mathrm{PG}}$) and normalize by the mean conductance. The resulting decay of the visibility $\mathcal{V}(T)$ with increasing temperature is shown in Figure 3D.

In the weak-backscattering limit, thermal averaging predicts a visibility $\mathcal{V}(T) = \mathcal{V}_A \frac{2L/L_T}{\sinh(2L/L_T)}$, with $L_T = \frac{hv}{2\pi^2 k_B T}$ being the thermal length[5]. Here, $L = 2\sqrt{A_{\mathrm{eff}}}$ is the one-way interfering length (Figure 1B), $\mathcal{V}_A$ is a size-dependent amplitude, $v$ is the edge velocity, and $k_B$ is the Boltzmann constant. A single velocity $v \approx (4.9 \pm 0.2) \times 10^4\ \mathrm{ms}^{-1}$ gives reasonable fits (solid curves in Figure 3D) to the three datasets over the measured temperature range, $T = 0.05 - 1.0$ K, demonstrating that the temperature dependence of the three cavity sizes is quantitatively described by a common thermal-averaging model. The extracted velocity at $B = 12$ T is approximately a factor of three lower than that reported for graphene QH FP interferometers[26,27], reflecting a smoother confinement potential for the edge channels and/or stronger screening effects in our device. The smoother confinement potential may come from the smaller interferometer sizes in our experiment; meanwhile, the substantially smaller cyclotron energy in monolayer $WSe_2$ likely gives stronger screening effects. More controlled experiments are required to better understand the underlying origin.

The size-dependent amplitude $\mathcal{V}_A = \mathcal{V}_0 \exp(-2L/L_\varphi)$ further provides an estimate of the phase-coherence length $L_\varphi$. Here, $\mathcal{V}_0 = 2\frac{\sqrt{(1-T_N)(1-T_S)}}{T_N+T_S-1} \approx 0.63$ is the ideal beam-splitter visibility set by the QPC transmissions, $T_N \approx 0.65$ and $T_S \approx 0.91$, and is independent of the cavity size (Supplementary Section S4). At the base temperature $T \approx 50$ mK, the thermal length $L_T \approx 2.4$ μm is much larger than the cavity perimeter, giving a negligible thermal reduction, i.e. $\frac{2L/L_T}{\sinh(2L/L_T)} \approx 0.98$. The reduction of the measured visibility below $\mathcal{V}_0$ is thus dominated by extrinsic dephasing rather than thermal averaging. Using the visibility measured at $T \approx 50$ mK together with the independently measured cavity perimeter $2L$ for each configuration, we obtain consistent values for $L_\varphi \approx 280$–$310$ nm (mean $\approx 290$ nm) (Methods).

The extracted coherence length is substantially shorter than the cavity perimeter, $2L \approx 740$–$880$ nm ($2L/L_\varphi \approx 2.4$–$3.1$), consistent with the observation of clear but much reduced interference visibility from the ideal value $\mathcal{V}_0$. Although the precise value of $L_\varphi$ may be subject to errors in the estimate of $\mathcal{V}_0$, the principal conclusion is robust: a single, size-independent coherence length, comparable to but smaller than the cavity perimeter, consistently describes the observations in all three interferometers.

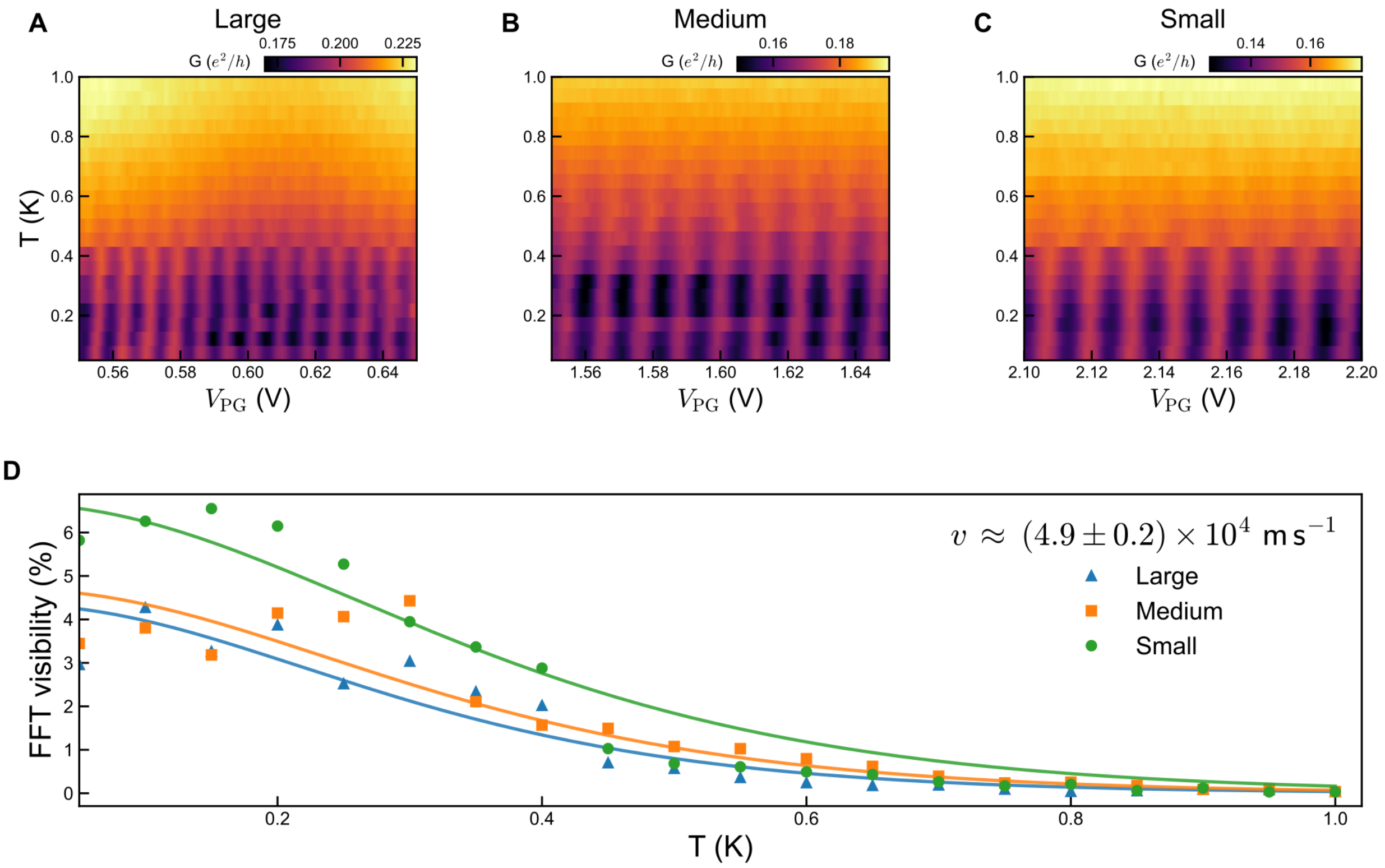

**Figure 3: Temperature dependence of the Aharonov–Bohm oscillations. A–C,** Diagonal conductance $G$ as a function of $V_{PG}$ and $T$ for the large, medium, and small cavity, respectively, measured at $B = 12$ T and bulk filling factor $\nu = -4$. **D,** Temperature dependence of the Fourier visibility $\mathcal{V}(T)$ for the three cavity sizes (markers). Solid curves are fits of the data to the thermal-averaging model (see main text). The fit yields an edge velocity $v \approx (4.9 \pm 0.2) \times 10^4\ ms^{-1}$.

## Bias dependence of the Aharonov–Bohm oscillations

Next, we study effects beyond the linear-response regime, where finite-bias measurements provide access to the characteristic energy and decoherence scales of the interfering edge states. Figures 4A, C, E show the diagonal conductance $G$ as a function of bias voltage drop across the interferometer ($V_{\mathrm{DC}}$) and $V_{\mathrm{PG}}$ for the three cavity sizes. $V_{\mathrm{DC}}$ is related to the applied DC bias voltage by a multiplication factor $f \approx 0.44$ that accounts for the series contact resistance in our device (Methods). In all three cases, the AB oscillations develop a characteristic tilted checkerboard pattern with increasing bias.

The corresponding Fourier visibility $\mathcal{V}(V_{\mathrm{DC}})$, obtained in the same way as in the temperature dependence study, exhibits a pronounced central lobe followed by rapidly decaying side lobes (Figures 4B, D, F), characteristic of finite-bias QH FP interference[5,44]. Notably, the visibility is strongly suppressed within approximately one central-lobe width. The central-lobe width $\Delta V$, defined as the separation between the first visibility minima on either side of zero bias, increases as the cavity size decreases, from $\Delta V = 277$ μV for the large cavity to $\Delta V = 296$ μV for the small cavity. The Fourier phase (bottom panels of Figures 4B, D, F) also changes by $\pi$ at the visibility minima. The $\pi$ phase slip correlates with the oscillations in $V_{\mathrm{DC}}$ at constant $V_{\mathrm{PG}}$. The results demonstrate bias-induced phase decoherence in the AB oscillations.

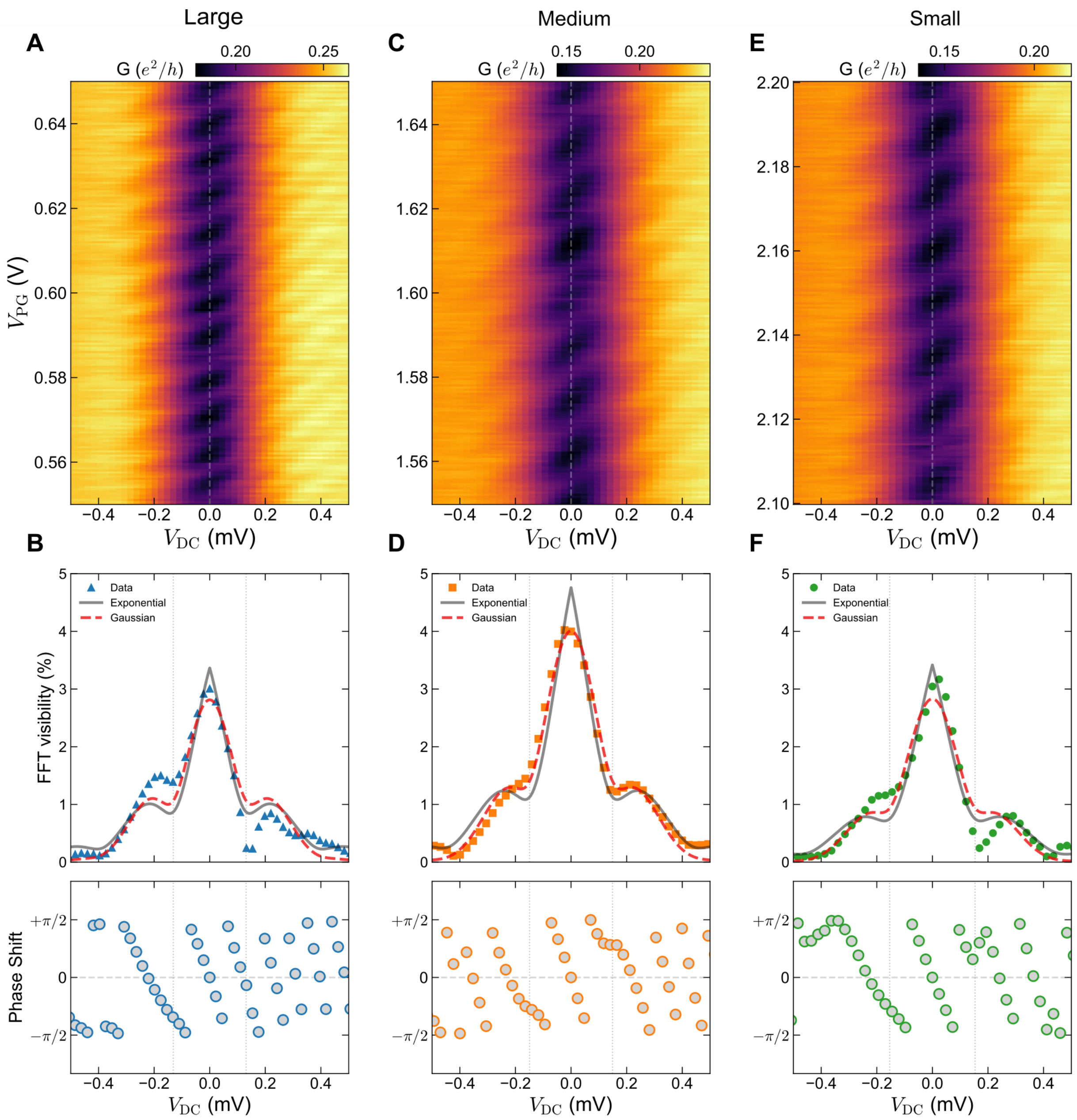

**Figure 4: Bias dependence of the AB oscillations. A, C, E,** Diagonal conductance $G$ as a function of the DC bias voltage $V_{DC}$ and $V_{PG}$ for the large, medium, and small configurations, respectively, showing a tilted checkerboard pattern for the AB oscillations. **B, D, F,** Fourier visibility (Top) and phase shift (referenced to $V_{DC} = 0$, Bottom) versus $V_{DC}$. Solid curves show fits to the data using exponential (solid green) and Gaussian (dashed purple) decoherence envelopes (Methods). $\pi$

phase slips are observed near the visibility nodes (vertical dotted lines). The fitted parameters are summarized in Table S1.

The observed bias dependence originates from the energy-dependent dynamical phase accumulated by electrons traversing the cavity. For a linearly dispersing edge state, $E = hvk/2\pi$, an energy shift $\delta E$ corresponds to a change in the wave vector by $\delta k = 2\pi\delta E/hv$ and hence the dynamical phase by $\delta k \times 2L$. Electrons within the finite source–drain energy window therefore acquire different dynamical phases, whose interference produces a periodic dependence on $V_{\mathrm{DC}}$ (Methods). The voltage period $\Delta V$ is set by the energy change $\Delta E = e\Delta V = \frac{hv}{2L}$ corresponding to a $2\pi$ dynamical phase shift (i.e. $\Delta k = \pi/L$)[5]. We can therefore independently determine an edge velocity $v \approx (5.6 \pm 0.3) \times 10^4\ \mathrm{ms^{-1}}$ from the bias dependence measurements. The value is in reasonable agreement with that determined from temperature dependence measurements. The agreement provides a self-consistency check and supports the association of the measured bias scale with the ballistic Thouless energy $E_{\mathrm{Th}} = \frac{hv}{L}$ rather than an energy scale set by disorder or charging.

The above picture also allows us to quantitatively model the bias dependence and reproduce the checkerboard pattern. By accounting for an asymmetric voltage drop across the two QPCs, a tilted checkerboard pattern with a periodic dependence on $V_{\mathrm{DC}}$ can be derived (Methods and Supplementary Figure S4). However, the simple picture based on non-interacting particles fails to capture the observed rapid decay of the visibility (Figures 4B, D, F). Similar rapid decay was observed in GaAs QH FP interferometers and attributed to interaction-induced dephasing[44]. Specifically, Coulomb interactions can produce an approximately exponential decay with bias[45], whereas Gaussian damping is associated with phase fluctuations of the interfering channels[3,4]. We compare our results to both the exponential and Gaussian decoherence envelopes (Methods and top panels in Figures 4B, D, F). Both describe the measurements comparably well; the present data quality is not able to distinguish the microscopic decoherence mechanisms.

## Conclusion

We have demonstrated coherent AB interference in a gate-defined FP interferometer formed from QH edge channels in monolayer $WSe_2$, establishing the first such device in a transition-metal dichalcogenide semiconductor. Despite the compact cavity size (with a perimeter of about 800 nm), we show that the device operates predominantly in the AB regime, consistent with a weak bulk–edge coupling estimated from the device electrostatics. By tuning a single device through three cavity sizes, we further show that the bias and temperature dependence of the interference are consistently described by a common ballistic picture, yielding an edge velocity $\approx 5 \times 10^4\ \mathrm{ms^{-1}}$, a ballistic Thouless energy ≈ 0.5 meV, and a phase-coherence length $\approx 300$ nm.

Beyond demonstrating coherent interferometry, monolayer $WSe_2$ offers a unique platform for QH physics: The strong spin–orbit coupling and spin–valley locking produce single-component Landau levels, avoiding the multi-component ambiguities present in graphene. Combined with the robust fractional QH states already observed in this material and its intrinsic optical addressability, our results establish monolayer $WSe_2$ as a promising platform for interferometric studies of single-component fractional QH states and anyonic statistics.

## Methods

### Device Fabrication

The quantum point contact (QPC) and plunger-gate electrodes were first patterned on a $Si/SiO_2$ substrate using electron-beam lithography, followed by electron-beam evaporation of Cr/Pd (2/10 nm). Few-layer graphite, multilayer hexagonal boron nitride (hBN) and monolayer $WSe_2$ flakes were mechanically exfoliated from bulk crystals onto $Si/SiO_2$ substrates and identified by their optical reflection contrast. The flakes were sequentially picked up using a polycarbonate film supported by a polydimethylsiloxane stamp in the following order: hBN, graphite top gate, hBN, monolayer $WSe_2$ and bottom hBN. The top and bottom hBN layers had thicknesses of 17 nm and 10 nm, respectively.

The assembled heterostructure was subsequently released at 200 °C onto a $Si/SiO_2$ substrate containing pre-patterned electrodes. Finally, the contact gates were fabricated using electron-beam lithography and electron-beam evaporation of Ti/Pd (5/40 nm). The graphite top gate defines the conducting channel, while the contact gates electrostatically induce high hole densities in the $WSe_2$ regions between the channel and the Pt electrodes, thereby reducing the contact resistance. An optical micrograph of the completed device is shown in Supplementary Figure S1.

### Electrical measurements

Measurements were performed in a Bluefors dilution refrigerator reaching a base electronic temperature of about 50 mK, equipped with a superconducting solenoid providing fields up to $B = 12$ T and copper-powder filters at the mixing chamber stage for electron thermalization.

Devices were measured in a four-terminal, voltage-bias configuration with an a.c. excitation of 50 µV using low-frequency lock-in amplifier techniques. The diagonal voltage $V_D$, taken between the $V_+$ and $V_-$ contacts (Figure 1C), was converted to the diagonal resistance $R_D = V/I$ and conductance $G = 1/R_D$ reported throughout. Current was measured using an Ithaco 1211 current amplifier and a Keithley 2400 source-measure unit. For DC bias measurements, the AC and DC voltages were combined at room temperature with a home-made passive adder circuit before being applied to the source while the drain was kept grounded through the current amplifier. The device-bias fraction $f \approx 0.44$ used to convert the applied voltage to the bias dropped across the interferometer was obtained from this configuration (see below).

### Thermal averaging and phase-coherence model

In the weak-backscattering limit, thermal averaging of the interference predicts a visibility[5]

$$\mathcal{V}(T) = \mathcal{V}_A \frac{2L/L_T}{\sinh(2L/L_T)}, \qquad L_T = \frac{hv}{2\pi^2 k_B T}, \tag{1}$$

where $v$ is the edge velocity, $L$ the one-way interfering length, $L_T$ the thermal length, and $\mathcal{V}_A$ a size-dependent amplitude. At high temperatures, Eq. (1) reduces to an exponential decay $\mathcal{V} \propto e^{-2L/L_T}$.

Writing the measured visibility as thermal averaging multiplied by an extrinsic decoherence factor gives

$$\mathcal{V}_A = \mathcal{V}_0 \, \exp\left(-\frac{2L}{L_\varphi}\right), \quad (2)$$

where $\mathcal{V}_0$ is the ideal beam-splitter contrast set by the QPC transmissions, and $L_\varphi$ is the phase-coherence length.

**Bias-divider calibration**

A DC voltage $V_{\mathrm{app}}$ is applied to the source contact, in series with which a finite contact resistance $R_s$ forms a voltage divider together with the interferometer resistance $R_{\mathrm{dev}}$. The two resistances are obtained from the measured two-probe and four-probe resistances as:

$$R_{\mathrm{dev}} = R_{4P}, \quad R_S = R_{2P} - R_{4P}$$

Near zero DC bias, we obtain $R_{4P} \approx 137$ kΩ and $R_{2P} \approx 312$ kΩ, corresponding to an effective series resistance $R_S \approx 175$ kΩ.

The voltage drop across the interferometer is therefore $V_{\mathrm{DC}} = f\, V_{\mathrm{app}}$, with the device-bias fraction

$$f = \frac{R_{\mathrm{dev}}}{R_{\mathrm{dev}} + R_S} \quad (3)$$

**Asymmetric finite-bias model**

For a linear edge dispersion $E = \hbar v k$, an energy shift gives a momentum change $\delta k = \delta E/\hbar v$, so one round trip of the loop with perimeter $2L$ accumulates a dynamical phase $(2L/\hbar v)\,\delta E$. Including an asymmetric bias drop, parameterized by a fraction $\beta = 1/2 + x$ at one QPC and $\bar{\beta} = 1/2 - x$ at the other ($x \in [-1/2, 1/2]$), the two branches acquire opposite bias-dependent offsets and the oscillatory conductance reads[27]

$$\delta G(V_{\mathrm{DC}}, \Phi) = \delta G_0 \left[\beta \cos\left(2\pi \frac{\Phi}{\Phi_0} - 2\pi \frac{2L}{hv}\, e\beta V_{\mathrm{DC}}\right) + \bar{\beta} \cos\left(2\pi \frac{\Phi}{\Phi_0} + 2\pi \frac{2L}{hv}\, e\bar{\beta} V_{\mathrm{DC}}\right)\right], \quad (4)$$

where $\Phi$ is the enclosed flux and $\Phi_0 = h/e$ the flux quantum (the effect of asymmetric voltage drop can be seen in Supplementary Figure S4).

The two cosines in Eq. (4) beat against one another and collapse into a single flux oscillation, $\delta G = \delta G_0\, A_{\mathrm{lobe}}(V_{\mathrm{DC}}) \cos[\, 2\pi\Phi/\Phi_0 - \psi(V_{\mathrm{DC}})\,]$, whose amplitude is the lobe envelope

$$A_{\mathrm{lobe}}(V_{\mathrm{DC}}) = \sqrt{\cos^2(2\pi\, eV_{DC}/E_{\mathrm{Th}}) + 4x^2 \sin^2(2\pi\, eV_{DC}/E_{\mathrm{Th}})}, \quad (5)$$

where$E_{\mathrm{Th}} = hv/L$ sets the characteristic bias scale. The first visibility minima occur at $V_{\mathrm{DC}} = \frac{\pm E_{\mathrm{Th}}}{4e}$, such that the central lobe width is $\Delta V = \frac{E_{\mathrm{Th}}}{2e}$. The phase offset $\psi(V_{\mathrm{DC}})$ tilts the checkerboard.

The experimental lobes decay much faster than Eq. (5) alone predicts. We therefore multiply the envelope by a phenomenological decoherence factor and compare the two forms established in the literature,

$$\text{exponential:} \quad \mathcal{V}(V_{\mathrm{DC}}) = A_0\, A_{\mathrm{lobe}}(V_{DC}) \exp(-2\pi\chi\, \mathrm{e}|V_{DC}|/E_{\mathrm{Th}}) \quad (6)$$

Gaussian: $\mathcal{V}(V_{\text{DC}}) = A_0\, A_{\text{lobe}}(V_{DC}) \exp(-V_{DC}^2/2V_0^2)$ (7)

with $A_0$ the zero-bias amplitude, $\chi$ a dimensionless damping rate, and $V_0$ a Gaussian width.

**Data availability**

The datasets generated and/or analyzed during the current study are available from the corresponding author upon reasonable request.

**Code availability**

The code used in this study is available from the corresponding author upon reasonable request.

**Acknowledgements**

H.C. thanks Preden Roulleau and C.L.T thanks Noah Samuelson and Liam Cohen for useful technical discussions. Funding for this work was provided in part by the KIC Fellowship and the Kavli Institute at Cornell for Nanoscale Science and by the Moore foundation, through OSP 146178. A portion of this work was performed at the National High Magnetic Field Laboratory, which is supported by National Science Foundation Cooperative Agreement No. DMR-2128556* and the State of Florida.

**Author contributions**

H.C, C.L.T, J.S and K.F.M conceived the experiment; C.L.T and J.F fabricated the device with the help of H.C. C.Y., S.C. and W.C. contributed to the early stages of this work. H.C and C.L.T. performed the measurements; H.C, C.L.T and S.B analyzed the data with inputs from J.S and K.F.M; T.T. and K.W. provided the hBN crystals.; All authors discussed the results and commented on the paper.

**Competing interests**

The authors declare no conflict of interest.

# Supplementary Information: Quantum Hall Fabry–Pérot Interferometry in Monolayer $WSe_2$

Himadri Chakraborti[1,2†*], Charles L. Tschirhart[2,3,4†*], James Fang[2], Saisab Bhowmik[1], Chengrui Yu[2], Shujian Cao[2], William Caro[2], Kenji Watanabe[5], Takashi Taniguchi[6], Jie Shan[1−4], Kin Fai Mak[1−4]

[1]Max Planck Institute for the Structure and Dynamics of Matter, Hamburg, Germany
[2]Laboratory of Atomic and Solid State Physics, Cornell University, Ithaca, NY 14853, USA
[3]School of Applied and Engineering Physics, Cornell University, Ithaca, NY 14853, USA
[4]Kavli Institute at Cornell for Nanoscale Science, Cornell University, Ithaca, NY 14853, USA
[5]Research Center for Electronic and Optical Materials, National Institute for Materials Science, Tsukuba, Japan
[6]Research Center for Materials Nanoarchitectonics, National Institute for Materials Science, Tsukuba, Japan
[†]These authors contributed equally.
[*]Correspondence: himadri.chakraborti@mpsd.mpg.de, cltschirhart@gmail.com

## S1 Device

An optical micrograph of an interferometer device is shown in Fig. S1.

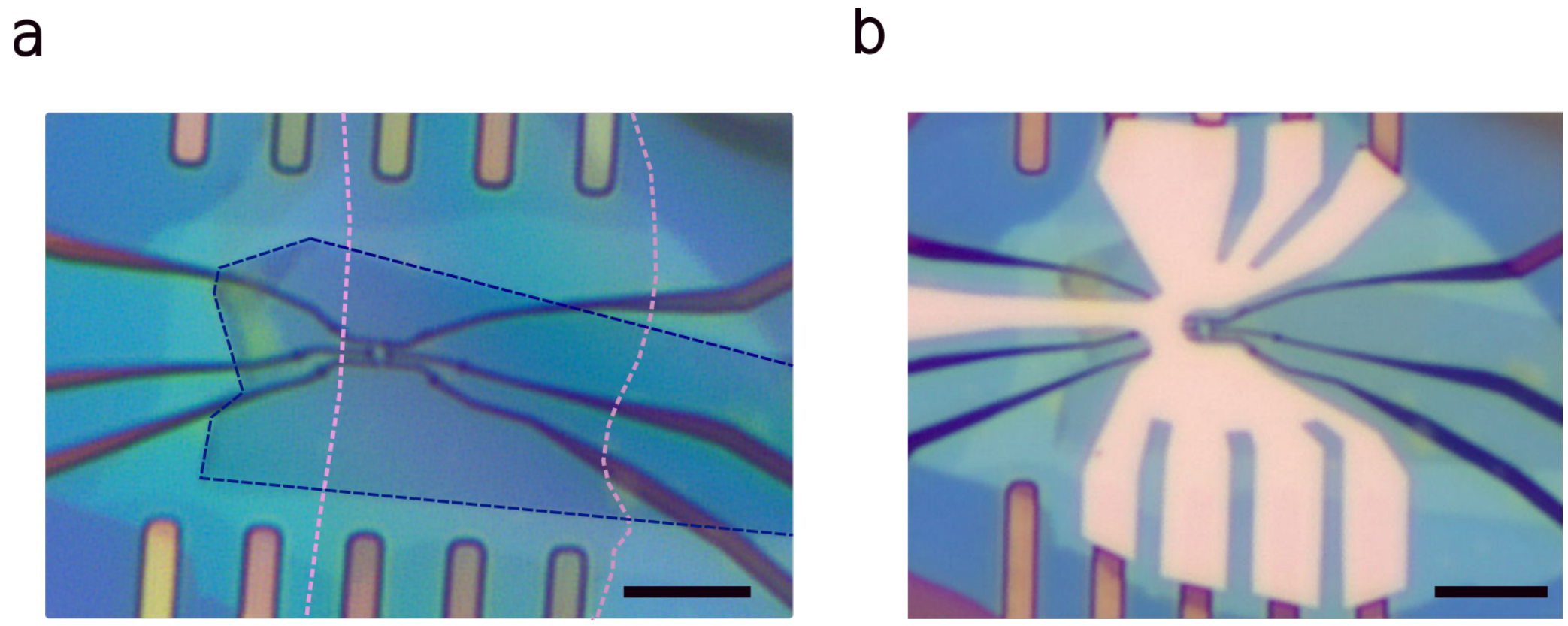

Figure S1: (a) Optical micrograph of an interferometer device. The monolayer $WSe_2$ flake and graphite top gate are outlined by light magenta and blue dotted lines, respectively. (b) Optical micrograph of the completed device after fabrication of the contact gates. Black scale bars in the lower-right corners of both panels represent 5 $\mu m$.

## S2 Electrostatic modulation of the QPC width

Figure S2a shows the four-probe conductance across the QPC, $G$, as a function of the QPC gate voltage, $V_{\mathrm{QPC}}$, measured at a fixed top-gate voltage of $V_{\mathrm{TG}} = -3$ V and $B = 0$ T. From the finite-bias spectroscopy of the QPC at $B = 0$ T, we estimate the gate lever arm, which relates changes in the split-gate voltage to the corresponding energy scale. As shown in Fig. S2b, the lever arm is extracted from the slope of the diamond-shaped feature surrounding the $G = 2e^2/h$ conductance plateau. The diamond boundaries are marked by yellow dashed lines, with one edge highlighted in cyan. From the slope of this edge, we obtain $\alpha = |\Delta V_{\mathrm{SD}}/\Delta V_{\mathrm{SG}}| \approx 5.8 \times 10^{-3}$. This lever arm allows us to convert the gate-voltage spacing between consecutive conductance plateaus into the corresponding subband energy spacing.

Using the extracted lever arm, we estimate the energy spacing between consecutive subbands, $\Delta E_{n,n+1}$, where $n$ denotes the subband index. Assuming a hard-wall confinement potential, the effective electrostatic width of the QPC can be estimated as

$$W_{\mathrm{eff}}(n) = \pi\hbar\sqrt{\frac{2n+1}{2m^*\Delta E_{n,n+1}}}, \tag{S1}$$

where $m^*$ is the effective hole mass in $WSe_2$. Unlike the lithographically defined width, $W_{\mathrm{eff}}$ is determined by the electrostatic confinement imposed by the split gates and can therefore be tuned by the applied gate voltages. As shown in Fig. S2c, the extracted effective width is approximately a factor of two smaller than the lithographically defined QPC width.

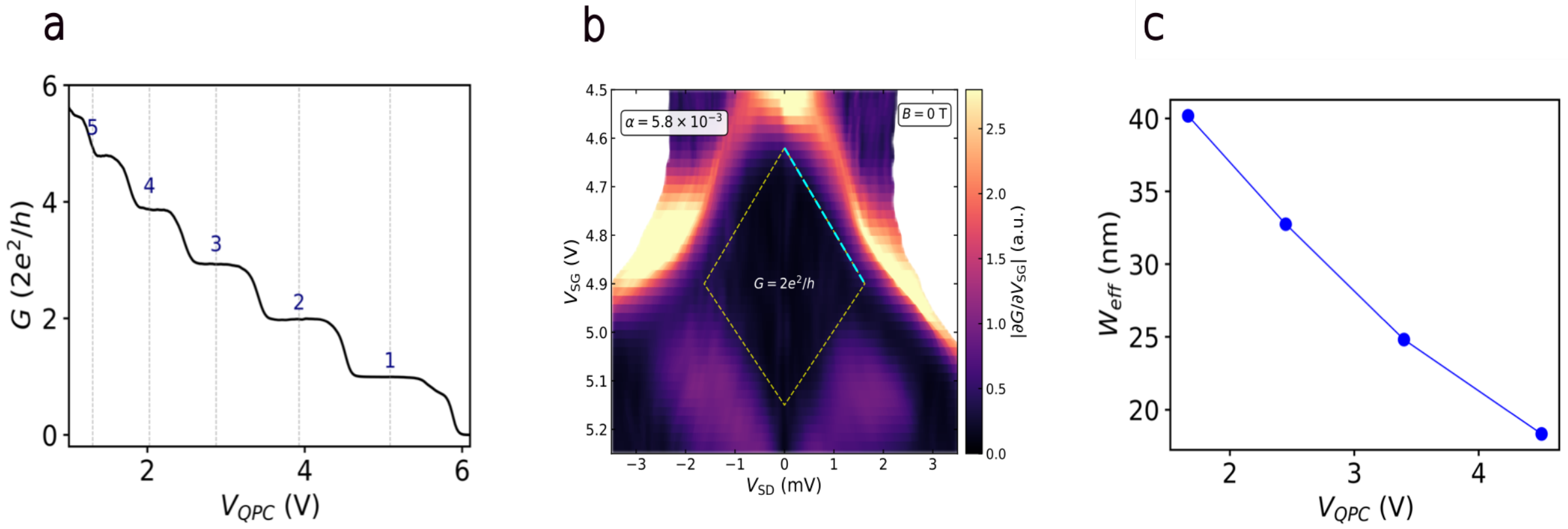

Figure S2: (a) Four-probe conductance $G$ as a function of the QPC voltage $V_{QPC}$, showing plateaus quantized in units of $2e^2/h$, measured at zero magnetic field ($B = 0$ T) and a temperature of $T = 50$ mK. (b) Finite-bias spectroscopy of the QPC at $B = 0$ T, showing the absolute conductance derivative, $|\partial G/\partial V_{SG}|$, as a function of source–drain bias $V_{SD}$ and split-gate voltage $V_{SG}$. The yellow dashed lines outline the diamond-shaped feature corresponding to the $G = 2e^2/h$ conductance plateau. The cyan dashed line highlights the diamond edge used to extract the gate lever arm, $\alpha = |\Delta V_{SD}/\Delta V_{SG}| \approx 5.8 \times 10^{-3}$. (c) Effective width of the QPC, $W_{eff}$, extracted from the subband spacing shown in (a), plotted as a function of $V_{QPC}$.

## S3 Bulk–Edge Coupling inside the interferometer cavity

The systematic phase jumps observed in the interference fringes (Main text Figures 2E-G) motivate a quantitative characterization of the relative contributions of the Aharonov–Bohm (AB) and Coulomb-dominated (CD) mechanisms to the interferometric phase. We estimate the dimensionless bulk–edge coupling parameter $\Delta = C_{IL}/(C_{IL} + C_L)$, where $C_{IL}$ is the edge–bulk capacitance and $C_L$ is the bulk capacitance.

For our device, consisting of a graphite top gate separated from the interferometer by an hBN dielectric of thickness $d_t = 17$ nm, the bulk capacitance is $C_L = (\epsilon_0 \epsilon_r / d_t) A_g$, where $\epsilon_0 = 8.854 \times 10^{-12}\ \mathrm{F\,m^{-1}}$ is the vacuum permittivity, $\epsilon_r \simeq 4$ is the dielectric constant of hBN, and $A_g = 0.048\ \mu\mathrm{m}^2$ is the geometric area enclosed by the interferometer. This yields $C_L \simeq 1.0 \times 10^{-16}$ F, corresponding to a charging energy $E_C = e^2/(2C_L) \simeq 0.8$ meV.

The edge–bulk capacitance is estimated using the model of Refs. [1, 2], in which the charge induced between the edge channels and the insulating bulk is distributed over a width set by the magnetic length, $l_B$. In this model, $C_{IL} = (2L\,\epsilon_0 \epsilon_r / 2\pi^2) \ln(4d_t/l_B)$, where $2L \simeq 880$ nm is the interferometer perimeter and $l_B = 7.4$ nm at $B = 12$ T. This gives $C_{IL} \simeq 3.5 \times 10^{-18}$ F.

Substituting these values into $\Delta = C_{IL}/(C_{IL} + C_L)$ gives $\Delta \simeq 0.034$. Since $C_{IL} \ll C_L$, the bulk–edge capacitive coupling is weak, implying that the interferometer area remains largely rigid and the device operates close to the Aharonov–Bohm limit. At the same time, the finite value of $\Delta$ indicates that Coulomb interactions are not entirely negligible. Notably, this estimate is comparable to the values $\Delta \sim 0.05$–$0.08$ reported for the first silicon-back-gated graphene Fabry–Pérot interferometers operating in the AB-dominated regime [1], suggesting a similar degree of bulk–edge decoupling despite the different material platforms.

## S4 QPC transmission and the ideal beam-splitter contrast $\mathcal{V}_0$

This section shows how we read off the two QPC transmissions and how we turned them into the ideal beam-splitter contrast $\mathcal{V}_0$ used in the main text.

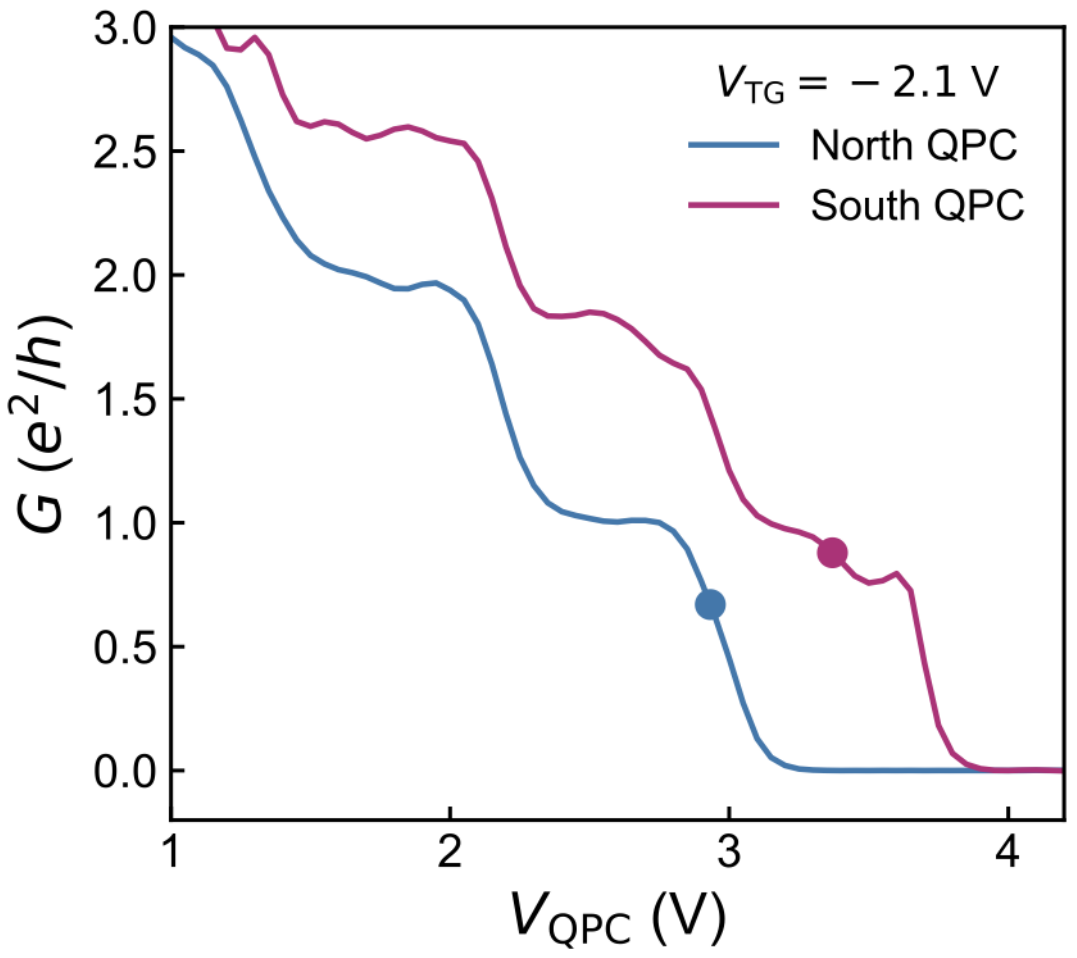

Figure S3: Individual transmission of south and north QPC at top gate voltage of $-2.1\ V$.

### S4.1 Extraction of $T_S$ and $T_N$

At the gate voltages used for interferometry, each QPC sits on its outermost, partially transmitted edge channel. All the channels below it are fully reflected, so none of them need to be added in separately. This means the conductance at that operating point already equals the transmission of the one channel that matters, in units of $e^2/h$,

$$T = \frac{G}{e^2/h}. \tag{S2}$$

Reading the conductance at the marked operating points in Fig. S3 gives

$$T_S \approx 0.91, \qquad T_N \approx 0.65, \tag{S3}$$

for the south and north QPC. We keep three significant figures here because $\mathcal{V}_0$ is sensitive to the difference between $T_S$ and $T_N$. The main text rounds these to $T_S \approx 0.91$ and $T_N \approx 0.65$ for readability. The correct value uses the unrounded transmissions in Eq (S3).

## S4.2 Ideal beam-splitter contrast

For two independent QPCs with transmissions $T_S$ and $T_N$, and reflection coefficients $R_S = 1 - T_S$ and $R_N = 1 - T_N$, the ideal visibility of a Fabry–Pérot interferometer (fully coherent, zero temperature) is

$$\mathcal{V}_0 = \frac{2\sqrt{(1-T_S)(1-T_N)}}{T_S+T_N-1}. \tag{S4}$$

With $T_S = 0.91$ and $T_N = 0.65$ this gives

$$\mathcal{V}_0 \approx 0.63. \tag{S5}$$

This is the contrast we use throughout the temperature and bias analysis.

# S5 Finite-bias Fourier visibility and extracted energy scales

To quantify the finite-bias evolution of the interference, we extract the Fourier amplitude of the Aharonov–Bohm (AB) oscillations at each dc bias. For each cavity, the dominant AB Fourier component is identified from the conductance oscillations near zero bias, and the same Fourier component is tracked as a function of $V_{\mathrm{DC}}$. The Fourier visibility is obtained by normalizing this amplitude to the mean conductance at each bias.

The resulting visibility is fitted using the asymmetric finite-bias lobe envelope derived above, multiplied by either an exponential or a Gaussian damping factor,

$$\mathcal{V}_{\exp}(V_{\mathrm{DC}}) = A_0 A_{\mathrm{lobe}}(V_{\mathrm{DC}}) \exp\left(-2\pi\chi \frac{e|V_{\mathrm{DC}}|}{E_{\mathrm{Th}}}\right), \tag{S6}$$

$$\mathcal{V}_{\mathrm{gau}}(V_{\mathrm{DC}}) = A_0 A_{\mathrm{lobe}}(V_{\mathrm{DC}}) \exp\left(-\frac{V_{\mathrm{DC}}^2}{2V_0^2}\right). \tag{S7}$$

Here $A_0$ is the zero-bias amplitude, $\chi$ is the dimensionless exponential damping parameter, and $V_0$ is the Gaussian decay scale.

The characteristic voltage period $\Delta V$ is set by the energy-dependent dynamical phase accumulated over the interfering path difference $2L$. For a linearly dispersing edge state, an energy shift $e\Delta V$ produces

$$\Delta\phi = \frac{2\pi e \Delta V}{hv}(2L). \tag{S8}$$

A phase change of $2\pi$ therefore corresponds to

$$e\Delta V = \frac{hv}{2L}. \quad \text{(S9)}$$

The corresponding ballistic Thouless energy is

$$E_{\mathrm{Th}} = \frac{hv}{L} = 2e\Delta V. \quad \text{(S10)}$$

The characteristic voltage scales obtained from the exponential and Gaussian fits are denoted by $\Delta V_{\mathrm{exp}}$ and $\Delta V_{\mathrm{gau}}$, respectively.

The one-way interfering length $L$ is independently estimated from the AB magnetic-field period $\Delta B$. Assuming one flux quantum $\Phi_0 = h/e$ threads the enclosed area gives $A = \Phi_0/\Delta B$. Approximating the cavity as a square of side length $a = \sqrt{A}$, the full interfering perimeter is $2L = 4a$, yielding $L = 2\sqrt{\Phi_0/\Delta B}$. The edge velocity and thermal scale are then obtained from

$$v = \frac{E_{\mathrm{Th}} L}{h}, \qquad T_0 = \frac{E_{\mathrm{Th}}}{4\pi^2 k_B}. \quad \text{(S11)}$$

Table S1 summarizes the geometric parameters, finite-bias fit parameters, and extracted energy scales for the three cavity sizes. The extracted edge velocities are of order $5 \times 10^4\ \mathrm{m\,s^{-1}}$, in reasonable agreement with the velocity independently obtained from the temperature dependence.

Table S1: Geometry, finite-bias fit parameters, and extracted quantities for the three cavity sizes. The characteristic voltage scales $\Delta V_{\mathrm{exp}}$ and $\Delta V_{\mathrm{gau}}$ are obtained from the exponential and Gaussian damping fits, respectively. The Thouless energy and edge velocity are calculated from $\Delta V_{\mathrm{exp}}$, using $E_{\mathrm{Th}} = 2e\Delta V_{\mathrm{exp}}$.

| | $\Delta B$ (mT) | $L$ ($\mu$m) | $\Delta V_{\mathrm{exp}}$ ($\mu$V) | $\chi$ | $\Delta V_{\mathrm{gau}}$ ($\mu$V) | $V_0$ ($\mu$V) | $E_{\mathrm{Th}}$ ($\mu$eV) | $v$ ($10^4\ \mathrm{m\,s^{-1}}$) | $T_0$ (mK) |
|---|---|---|---|---|---|---|---|---|---|
| Large | 85.7 | 0.439 | 277 | 0.42 | 289 | 183 | 554 | 5.88 | 163 |
| Medium | 100 | 0.407 | 295 | 0.48 | 311 | 173 | 589 | 5.80 | 173 |
| Small | 120 | 0.371 | 296 | 0.53 | 302 | 162 | 591 | 5.30 | 174 |

## S6 Finite-bias tilted checkerboard

The finite-bias conductance map exhibits a tilted Aharonov–Bohm checkerboard pattern, indicating an asymmetric voltage distribution between the two QPCs (Supplementary Fig. S4a). To reproduce this behavior, we simulate the conductance using the asymmetric finite-bias interference model described in Eq. (4) of the main text. Using an asymmetry parameter $x = 0.25$, an edge velocity $v = 5.88 \times 10^4\ \mathrm{m\,s^{-1}}$, and a Gaussian damping scale $V_0 = 183\,\mu\mathrm{V}$, the simulation reproduces both the tilt of the interference fringes and their suppression at finite bias (Supplementary Fig. S4b).

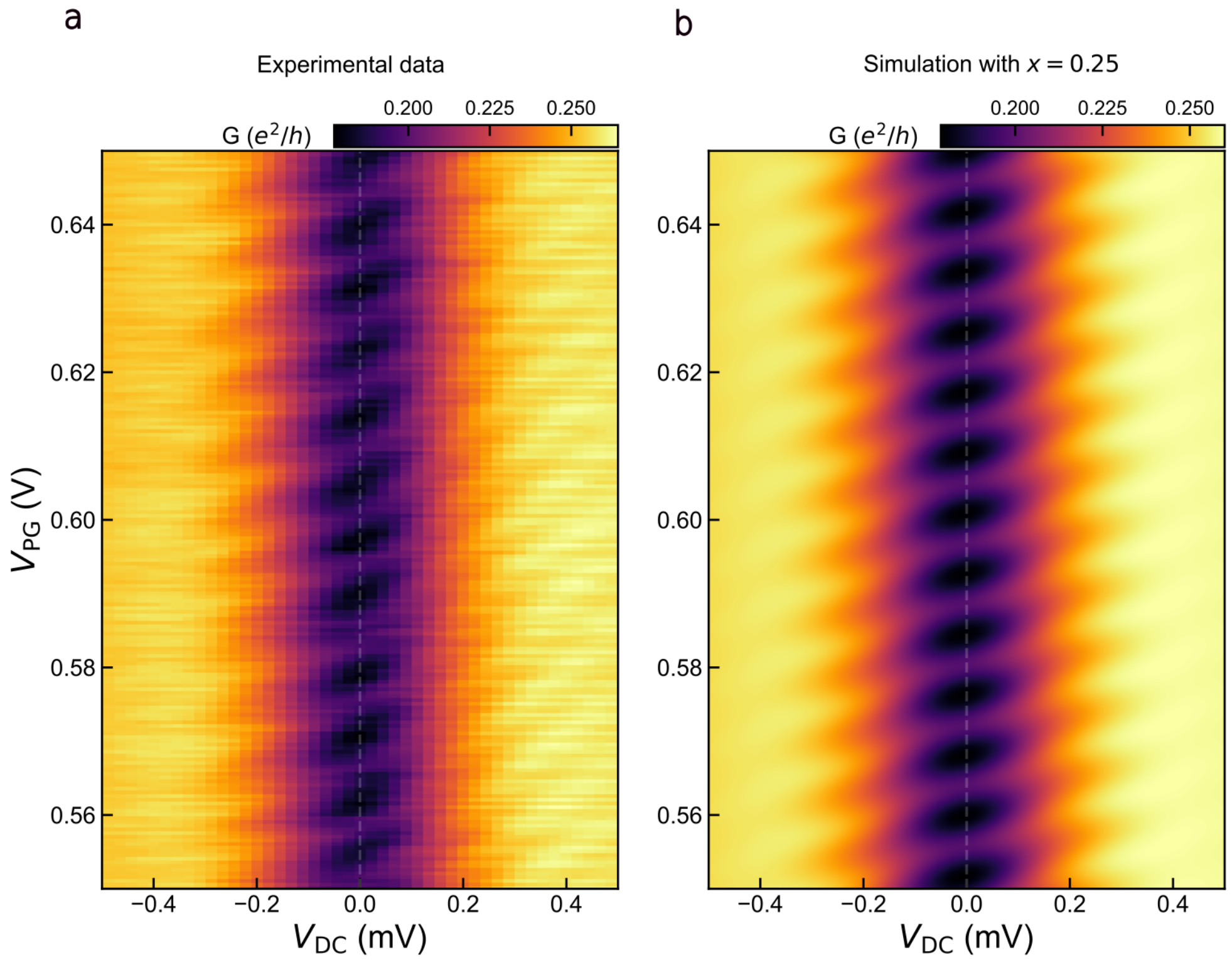

Figure S4: **Bias-dependent asymmetry of the Aharonov–Bohm checkerboard. (a)** Experimental diagonal conductance $G$ as a function of dc bias $V_{DC}$ and plunger-gate voltage $V_{PG}$ for the large cavity, measured at $B = 12\,T$, $T = 50\,mK$, and bulk filling factor $\nu = -4$. The Aharonov–Bohm fringes form a tilted checkerboard pattern that is visible only within a narrow window around $V_{DC} = 0$ (dashed line) and washes out into a smooth, gate-independent background by $|V_{DC}| \gtrsim 0.3$ mV. **(b)** Simulated conductance map for the same cavity, computed from the asymmetric finite-bias model [Eq. 4, with bias-asymmetry parameter $x = 0.25$, one-way length $L = 0.439\,\mu m$, and edge velocity $v = 5.88 \times 10^4\,m\,s^{-1}$ (Table S1, large cavity) and Gaussian decoherence width $V_0 = 183\,\mu V$. To enable a direct comparison with the raw conductance in **(a)**, the smooth bias-dependent background was taken directly from the data, the column-wise mean $\langle G\rangle(V_{DC})$ of the experimental map and added to the simulated oscillatory term $\delta G(V_{DC}, V_{PG})$. The simulation reproduces both the tilt direction and the bias range over which fringes remain visible, confirming that an asymmetric voltage drop with $x = 0.25$ between the two QPCs accounts for the skew of the checkerboard pattern observed experimentally.

## S7 Validity regime of the exponential coherence factor

Figure S5 plots the extrinsic coherence factor $\exp(-2L/L_\varphi)$ using the phase-coherence length $L_\varphi \approx 0.29$ µm obtained in the main text. For the three cavities studied here, this gives $2L/L_\varphi \approx$ 2.6 to 3.0, so all three sit in the regime $2L \gtrsim L_\varphi$. This is the regime where the exponential form is the right description of the coherence factor, which is why we use it in the visibility analysis. The corresponding surviving contrast is $\exp(-2L/L_\varphi) \approx 0.05$ to 0.08, consistent with interference that is clearly visible but substantially reduced.

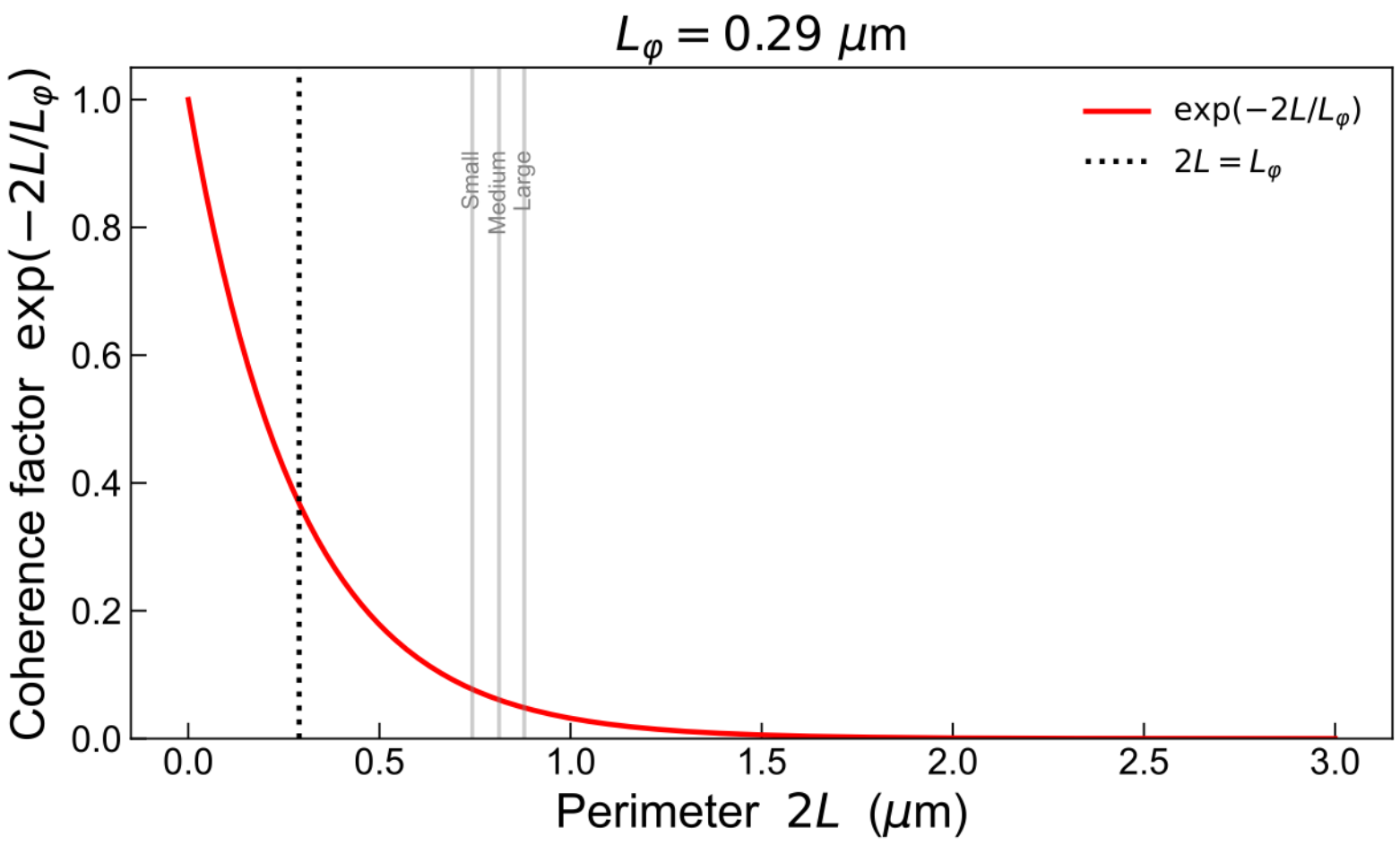

Figure S5: **Validity regime of the exponential coherence factor.** Coherence factor $exp(-2L/L_\varphi)$ (red) versus the interferometer perimeter $2L$, evaluated for the extracted phase-coherence length $L_\varphi = 0.29$ µm. This factor is the extrinsic-dephasing term of the visibility model (equation (2) in methods of main text), which multiplies the thermal-averaging contribution; it quantifies the fraction of interference contrast that survives dephasing after an electron traverses the cavity once. The dotted vertical line marks $2L = L_\varphi$, where the coherence factor equals $e^{-1} \approx 0.37$. The three gray vertical lines indicate the perimeters of the large, medium, and small cavities studied here ($2L = 0.88$, $0.81$, and $0.74$ µm), which all lie to the right of the $2L = L_\varphi$ line, at $2L/L_\varphi \approx$ 2.6–3.0. In this regime ($2L \gtrsim L_\varphi$) the exponential decay is the appropriate description of the coherence factor, justifying its use in the visibility analysis; the corresponding surviving contrast is $exp(-2L/L_\varphi) \approx 0.05 - 0.08$, consistent with the clear but substantially reduced interference observed. For $2L \lesssim L_\varphi$ (left of the dotted line) the curve instead rises steeply toward unity and the simple exponential overestimates the dephasing, since in this short-path limit an electron acquires little phase uncertainty per traversal and the true coherence factor saturates below one. The narrow spread of perimeters ($\sim$18% in $2L$) places all three cavities at nearly the same point on the curve, which is why a single coherence length $L_\varphi \approx 0.29$ µm is quoted for the device rather than three independently resolved values; distinguishing a size dependence of $L_\varphi$ would require a substantially wider range of $2L$.